\documentclass[11pt,a4paper]{article}

\usepackage[T1]{fontenc}
\usepackage[utf8]{inputenc}
\usepackage{newtxtext}
\usepackage{amsmath,amssymb}
\usepackage{newtxmath}
\usepackage{microtype}

\usepackage{geometry}
\usepackage{setspace}
\usepackage{xcolor}
\usepackage{graphicx}
\usepackage{caption}
\usepackage[FIGTOPCAP]{subfigure}
\usepackage{siunitx}
\usepackage{enumerate}
\usepackage{booktabs}
\usepackage{multirow}
\usepackage{tabularx}
\usepackage{array}
\usepackage{titlecaps}
\usepackage{balance}
\usepackage{footmisc}
\usepackage{lineno}
\usepackage{titlesec}
\usepackage{dirtytalk}
\usepackage{tikz}
\usetikzlibrary{positioning}
\usepackage[disable,textsize=footnotesize]{todonotes}

\definecolor{headingcolor}{HTML}{000000}
\definecolor{bodygray}{HTML}{333333}
\definecolor{mutedgray}{HTML}{626B73}
\definecolor{linkcolor}{HTML}{245D87}
\definecolor{rulecolor}{HTML}{D5DCE2}
\definecolor{quotecolor}{HTML}{333333}
\definecolor{correct}{HTML}{5B8E55}
\definecolor{incorrect}{HTML}{B8C0C8}
\definecolor{textgray}{HTML}{333333}
\definecolor{wcblue}{HTML}{4C78A8}
\definecolor{wcgreen}{HTML}{59A14F}
\definecolor{wcorange}{HTML}{F28E2B}
\definecolor{wcpurple}{HTML}{9C6DB2}
\definecolor{wcred}{HTML}{E15759}
\definecolor{wcteal}{HTML}{4EAAA0}

\usepackage[
    backend=biber,
    style=numeric,
    sorting=nyt,
    maxnames=3,
    minnames=1
]{biblatex}
\usepackage{xurl}
\usepackage{hyperref}
\usepackage{orcidlink}
\hypersetup{
    colorlinks=true,
    linkcolor=black,
    citecolor=linkcolor,
    urlcolor=linkcolor,
    pdfborder={0 0 0},
    pdftitle={Between Obligations and Evidence: A Participatory Study of Translating the EU AI Act into Technical Requirements},
    pdfauthor={Helma Torkamaan, Masoud Mansoury, Yanan Xin, Oscar Oviedo Trespalacios, Roberta Savella, Andrej Zwitter, Maria Heuss, Huy Son Nguyen, Marie-Therese Sekwenz, Moyi Li, Roberta Tamponi, Megha Khosla}
}

\usepackage{fancyhdr}
\newcommand{\shorttitle}{Between Obligations and Evidence}
\newcommand{\shortauthors}{Torkamaan et al.}
\fancypagestyle{firstpage}{%
    \fancyhf{}%
    \fancyfoot[C]{\small\textcolor{mutedgray}{\thepage}}%
}

\titleformat{\section}
    {\large\bfseries\color{black}}
    {\thesection}{0.65em}{}
\titleformat{\subsection}
    {\normalsize\bfseries\color{black}}
    {\thesubsection}{0.65em}{}
\titleformat{\subsubsection}[runin]
    {\normalsize\bfseries\color{black}}
    {\thesubsubsection}{0.6em}{}[.]

\titlespacing*{\section}{0pt}{15pt plus 2pt minus 2pt}{5pt}
\titlespacing*{\subsection}{0pt}{11pt plus 2pt minus 1pt}{4pt}
\titlespacing*{\subsubsection}{0pt}{7pt plus 2pt minus 1pt}{0.5em}

\let\oldsay\say
\renewcommand{\say}[1]{%
    \oldsay{\textcolor{quotecolor}{\textit{#1}}}%
}

\newcommand{\mpar}[1]{\textbf{#1}.}

\newlength{\optionwidth}
\newlength{\barwidth}
\title{Between Obligations and Evidence: A Participatory Study
of Translating the EU AI Act into Technical Requirements}
\author{Helma Torkamaan, Masoud Mansoury, Yanan Xin,
Oscar Oviedo Trespalacios, Roberta Savella, Andrej Zwitter,
Maria Heuss, Huy Son Nguyen, Marie-Therese Sekwenz, Moyi Li,
Roberta Tamponi, Megha Khosla}
\date{}

\newcommand{\paperperson}[3]{%
    \mbox{#1\textsuperscript{#2}%
        \if\relax\detokenize{#3}\relax
        \else\,\orcidlink{#3}\fi}%
}
\newcommand{\papersep}{\hspace{0.8em}}
\newcommand{\paperline}[1]{\noindent#1\par}
\newcommand{\authoremail}[1]{\href{mailto:#1}{#1}}

\makeatletter
\renewcommand{\maketitle}{%
    \thispagestyle{firstpage}%
    \begingroup
    \centering
    \setstretch{1.0}%
    \setlength{\parskip}{0pt}%
    \setlength{\parindent}{0pt}%

    {\fontsize{19}{22}\selectfont\bfseries\color{black}
        \@title\par}
    \vspace{8pt}

    {\fontsize{10.5}{13}\selectfont
        \paperline{%
            \paperperson{Helma Torkamaan}{1}{0000-0003-1094-4059}%
            \papersep
            \paperperson{Masoud Mansoury}{1}{0000-0002-9938-0212}%
            \papersep
            \paperperson{Yanan Xin}{1}{0000-0003-3866-821X}%
            \papersep
            \paperperson{Oscar Oviedo Trespalacios}{1}{0000-0001-5916-3996}}
        \paperline{%
            \paperperson{Roberta Savella}{2,3}{0009-0007-3005-4252}%
            \papersep
            \paperperson{Andrej Zwitter}{4}{0000-0001-7391-141X}%
            \papersep
            \paperperson{Maria Heuss}{5}{0000-0001-5360-9627}%
            \papersep
            \paperperson{Huy Son Nguyen}{1}{0009-0006-4616-0976}}
        \paperline{%
            \paperperson{Marie-Therese Sekwenz}{1}{0000-0002-3686-6100}%
            \papersep
            \paperperson{Moyi Li}{1}{0009-0008-9638-7890}%
            \papersep
            \paperperson{Roberta Tamponi}{6}{}}
        \paperline{\paperperson{Megha Khosla}{1}{0000-0002-0319-3181}}
    }
    \vspace{2pt}

    {\fontsize{8.5}{10.5}\selectfont
        \authoremail{H.Torkamaan@tudelft.nl}\quad
        \authoremail{M.Mansoury@tudelft.nl}\quad
        \authoremail{Yanan.Xin@tudelft.nl}\par
        \authoremail{O.OviedoTrespalacios@tudelft.nl}\quad
        \authoremail{roberta.savella@isti.cnr.it}\quad
        \authoremail{andrej.zwitter@aau.at}\par
        \authoremail{m.c.heuss@uva.nl}\quad
        \authoremail{H.S.Nguyen@tudelft.nl}\quad
        \authoremail{M.T.Sekwenz@tudelft.nl}\par
        \authoremail{r.tamponi@liacs.leidenuniv.nl}\par
        \authoremail{M.Khosla@tudelft.nl}\par
    }
    \vspace{4pt}

    {\fontsize{9}{11}\selectfont\color{bodygray}
        \textsuperscript{1}Delft University of Technology,
        Delft, The Netherlands\par
        \textsuperscript{2}Institute of Information Science and Technologies
        (ISTI), National Research Council (CNR), Pisa, Italy\par
        \textsuperscript{3}University of Pisa, Pisa, Italy\quad
        \textsuperscript{4}University of Klagenfurt, Klagenfurt, Austria\par
        \textsuperscript{5}University of Amsterdam,
        Amsterdam, The Netherlands\par
        \textsuperscript{6}Leiden Institute of Advanced Computer Science
        (LIACS), Leiden University, Leiden, The Netherlands\par
    }
    \endgroup
    \par\vspace{11pt}
}
\makeatother

\renewenvironment{abstract}{%
    \par\noindent
    \begin{minipage}{\textwidth}%
    \setstretch{1.04}%
    \setlength{\parskip}{3pt}%
    \small\noindent\textbf{Abstract.}\enspace\ignorespaces
}{%
    \par\end{minipage}\par\vspace{4pt}%
}
\newcommand{\paperkeywords}[1]{%
    \par\noindent{\small\textbf{Keywords:} #1\par}%
    \vspace{4pt}%
}

\begin{document}

\maketitle


\begin{abstract}
The EU AI Act imposes technical and organizational requirements for high-risk AI systems, but applying these requirements involves contextual choices about what to measure and what evidence supports a compliance claim. We investigate these challenges through a participatory workshop involving 32 researchers, policymakers, and practitioners from different disciplines, using realistic scenarios in healthcare, recruitment, and public transport to examine how regulatory obligations are interpreted, translated into technical requirements, and evaluated in practice. Through a thematic analysis of workshop data, we identify recurring tensions between how regulatory obligations are translated into measurable system properties, how responsibility and human oversight are exercised in practice, and how the validity of compliance evidence is affected by changing operational conditions and system uses. For example, participants highlighted how fairness metrics could conceal discrimination in recruitment, how time pressure and limited information could undermine human oversight in public transport, and how changes in patient populations could invalidate previously established evidence of system performance. Our findings suggest that technical, organizational, and evidentiary challenges are interdependent, and that establishing and maintaining compliance requires attention not only to technical assessments but also to how AI systems are used, governed, and monitored over time. We discuss implications for supporting context-sensitive and ongoing compliance.
\end{abstract}

\paperkeywords{EU AI Act, technical requirements,
participatory research, AI governance, regulatory compliance, workshop}

\section{Introduction}

The European Union's Artificial Intelligence Act (AI Act)~\cite{EUAIAct2024} is a horizontal regulatory framework that governs AI systems through a risk-based logic. Systems classified as high-risk are subject to a package of technical and governance obligations, including risk management, documentation, oversight, and lifecycle monitoring. In addition, a parallel classifier for general-purpose AI models identifies a systemic risk threshold that imposes additional obligations. In this sense, the Act shifts AI governance from voluntary ethics-by-design toward compliance-by-design: providers and deployers must produce evidence that obligations are met, not merely state design principles.

A central challenge is translating these legal obligations into concrete technical requirements. Provisions concerning data governance (Art. 10), transparency (Art. 13), human oversight (Art. 14), and accuracy, robustness, and cybersecurity (Art. 15) do not directly specify which system properties engineers should implement, which metrics or tests they should use, or what evidence should demonstrate that a requirement has been satisfied. The AI Act anticipates parts of this implementation problem to be solved through instruments intended to support compliance. For example, harmonized standards (Art. 40), common specifications (Art. 41), and codes of practice (Art. 56) ~\cite{EUAIAct2024, gpaicop2025}. However, these instruments do not eliminate the need to interpret legal requirements within specific technical and organizational contexts \cite{gstreinzwitter2024AIA}. These decisions cannot be made by developers alone. They depend on how AI systems are integrated into organizational workflows, how responsibilities are distributed, and how people affected by these systems are considered in design and evaluation.

The challenge extends beyond specifying technical requirements. An AI system may satisfy predefined performance criteria without adequately supporting the decisions for which it is used. Similarly, providing human oversight mechanisms does not guarantee that people have the information, competence, or authority to intervene effectively. Compliance also depends on organizational arrangements and operating conditions that may change after deployment~\cite{torkamaan_role_2024}. These challenges raise questions about whether technical evidence alone can adequately demonstrate that regulatory obligations are being met, and how such evidence should be interpreted and maintained across different contexts of use.

While much of the current focus is on how regulatory obligations can be translated into technical requirements, evaluation metrics, and documentation, a distinct challenge is understanding whether these measures remain effective once embedded in organizational responsibilities and real deployment settings. In particular, there is a need to understand where translating legal requirements into technical and organizational arrangements becomes contested, and what this means for establishing credible evidence of compliance. A key unresolved issue is therefore not simply how to measure whether an AI system meets a technical requirement, but how to determine whether that measurement provides sufficient evidence that the system can be used responsibly in a particular context.

This paper examines the challenges of translating AI Act obligations into technical requirements, organizational practices, and evidence of compliance. We draw on SIG AI ACT, an interdisciplinary special interest group bringing together expertise in AI development, deployment, law, and policy. Through a participatory workshop involving 32 researchers and practitioners (Section~\ref{methods}), we investigate where this translation becomes difficult, how technical, organizational, and evidentiary challenges intersect, and what may help address these challenges. Our investigation is guided by three research questions covering the technical, sociotechnical, and legal dimensions of AI Act implementation:

\begin{description}
    \item[RQ1 -- Technical translation:] What challenges arise when translating AI Act requirements into technical requirements for AI systems?

    \item[RQ2 -- Sociotechnical translation:] What challenges arise when AI Act requirements are integrated into human, organizational, and deployment practices?

    \item[RQ3 -- Legal--evidentiary translation:] What challenges arise when interpreting AI Act requirements in their legal context and determining what evidence is sufficient to demonstrate their satisfaction?
\end{description} 
 \section{Related Work}

The EU AI Act (Regulation (EU) 2024/1689)~\cite{EUAIAct2024} regulates AI systems according to the risk they pose and seeks to balance AI innovation with the protection of fundamental rights, health, and safety. Under Article 6, high-risk AI systems include those used as safety components of certain regulated products and those operating in areas listed in Annex III, such as employment, education, law enforcement, migration, and essential services. These systems are subject to requirements (Art.~9--15) concerning risk management, data governance, transparency, human oversight, and robustness. Their providers should demonstrate satisfaction of these requirements through conformity assessment (Art.~43), which for most Annex~III systems takes the form of internal control (Annex~VI) and rests primarily on technical documentation (Art.~11).

The requirements themselves are formulated at different levels of specificity. For high-risk AI systems, the Act uses context-dependent formulations, requiring ``appropriate'' accuracy (Art.~15), ``sufficiently representative'' data (Art.~10), and effective human oversight (Art.~14)~\cite{EUAIAct2024}. These requirements need further specification to guide implementation and evaluation for particular systems~\cite{nolte_robustness_2025, marin2026your, kilian_european_2025, langer2025complexities}. Even where the Act is precise, such as in using the amount of computational resources used for training as a threshold for identifying general-purpose models with systemic risk, it still relies on the loosely defined notion of ``high-impact capabilities'' (Art.~51)~\cite{gstreinzwitter2024AIA}. Such measurable criteria capture only some aspects of a model's capabilities and cannot, on their own, establish whether broader regulatory requirements are satisfied.
Harmonized standards are expected to supply further specification, but legal analyses question how far standards can settle normative questions concerning fundamental rights~\cite{gornet2024european}, and public guidance helps organizations identify applicable obligations rather than implement them~\cite{ec2025servicedesk, bnetza2025kompass}.

\mpar{From obligations to technical requirements and evidence}
Translating regulation into software requirements is a long-standing concern in requirements engineering, not a challenge created by the AI Act alone~\cite{kosenkov2025systematic, otto_addressing_2007}. Prior work shows that deriving requirements from regulation requires statement-level analysis, cross-reference resolution, ambiguity handling, and prioritization where rights and obligations conflict~\cite{breaux2008analyzing}. Similar efforts have been developed for the GDPR~\cite{ayala2018grace} and, more recently, with LLM support for extracting requirements and tracing regulatory change~\cite{abualhaija2025llm, abualhaija2024ai}.
For the AI Act, COMPL-AI\footnote{https://compl-ai.org/} maps requirements onto existing LLM benchmarks and finds shortcomings in both model performance and the available benchmarks~\cite{guldimann2024compl}.
Such mappings show that obligations can be decomposed into testable properties. They also show that the decomposition remains partial.

A second source of difficulty lies in the translation from normative concepts to technical properties. Research on sociotechnical abstraction, problem formulation, and measurement validity shows that translating normative concepts into technical properties involves contestable choices and contextual assumptions \cite{selbst_fairness_2019, passi_problem_2019, jacobs_measurement_2021}. Performance metrics alone establish neither a system’s fitness for purpose nor the legitimacy of its decision procedure \cite{raji_fallacy_2022, wang2024against}. Under the AI Act, assessing “appropriate” accuracy requires choices about metrics, evaluation data, and acceptance thresholds \cite{marin2026your}. In EU non-discrimination law, fairness metrics cannot replace contextual legal judgment, and choosing metrics that preserve or transform existing inequalities has legal implications \cite{wachter_why_2021, weerts_algorithmic_2023, wachter_bias_2021}. Selecting a metric or test therefore does not establish whether the evidence suffices for an obligation, decision, or deployment context. This distinction becomes especially important when compliance depends on organizational practices and human judgment that cannot be assessed through technical testing alone.

Documentation approaches provide structured formats for recording and communicating information that can support compliance assessment, including model cards~\cite{mitchell_model_2019}, datasheets~\cite{gebru_datasheets_2021}, FactSheets~\cite{arnold_factsheets_2019}, and Data Cards~\cite{pushkarna_data_2022}, which predate the AI Act (for an overview, see~\cite{micheli_landscape_2023}).
In the European regulatory context, research has examined how documentation approaches address high-risk AI requirements~\cite{hupont_documenting_2023} and proposed use case cards~\cite{hupont2024use}, AI Cards~\cite{golpayegani2024ai}, TechOps templates~\cite{lucaj_techops_2025}, and the Model Documentation Form of the General-purpose AI (GPAI) Code of Practice~\cite{gpaicop2025}.
Existing Machine Learning Operations (MLOps) tools, such as model registries, dataset versioning, logging, and monitoring, can supply the underlying records, although they were designed for engineering rather than legal accountability~\cite{kreuzberger2023MLOPS}.
These approaches support transparency and traceability.
However, documentation alone does not necessarily establish whether regulatory requirements have been satisfied, particularly when compliance depends on specific deployment conditions and organizational practices~\cite{cobbe_reviewable_2021, sabuncuoglu_justified_2025, kuehnert_disclosure_2026}.
Complementary work on third-party assurance combines stakeholder interviews, artifact review, and a maturity matrix to assess development and deployment processes alongside system outcomes, broadening assessment to include organizational practices~\cite{kim_toward_2026}.

\mpar{Oversight, responsibility, and organizational capacity}
The Act's human oversight requirement has drawn sustained critique. Existing works argue that oversight policies tend to proceduralize discretion without ensuring that the overseer can exercise it~\cite{koulu_proceduralizing_2020}, that people may be unable to perform the supervisory role such policies assume~\cite{green_flaws_2022, lauxInstitutionalisedDistrustHuman2024}, and that effective oversight requires multiple conditions to be met jointly
~\cite{sterz2024quest}. Responsibility raises a parallel problem. The Act assigns obligations to providers and deployers, but AI systems are built and operated through chains of actors in which no single party controls data, model, and use~\cite{nissenbaum1997accountability, cobbe2023understanding}. Accountability in such settings is fundamentally relational~\cite{cooper2022accountability}. 
AI engineers may recognize harms yet see addressing them as beyond their responsibility or capacity~\cite{widder_dislocated_2023}.
Article~27's fundamental rights impact assessment leaves coordination between providers and deployers insufficiently specified~\cite{mantelero2024fundamental}.
Work on safety-critical technologies likewise identifies tensions between responsibility, authority, and the capacity to maintain or withdraw a technology once operational dependencies have formed~\cite{oviedotrespalacios2026holding}.
Article~4 also requires providers and deployers to take measures to support AI literacy among relevant personnel~\cite{EUAIAct2024}. Surveys document AI training provision~\cite{bitkom2025, bitkom2026} and work on AI governance alongside anticipated needs for additional governance staff~\cite{iapp2025aigov}. Recognizing an obligation thus differs from having the expertise, information, resources, and authority to discharge it.

\mpar{Life-cycle dependencies and sociotechnical aspects}
Conformity assessment establishes compliance at a point in time, but the evidence supporting it can lose validity through model updates, drift in data or populations, and changes in deployment conditions~\cite{mokander_conformity_2022, gotsch_harmonising_2024}. Dataset shift can undermine clinical AI performance~\cite{finlayson_clinician_2021}, while function creep can extend systems beyond their original purposes~\cite{koops_concept_2021}. Proposed approaches address reassessment through lifecycle models, process-based compliance methodologies, and continuous auditing \cite{gotsch_harmonising_2024, amanteaMethodologyComplianceAI2026, knoblauchContinuousAuditingBased2025}.
Applying these reassessment approaches also requires attention to how safeguards function within organizational workflows.
Research on the integration of LLMs into sociotechnical systems highlights the importance of human-centered design, longitudinal evaluation, and coordination across technical and organizational actors \cite{torkamaan_challenges_2024}. 
These perspectives suggest that sustaining compliance assurance requires evidence both that technical assessments remain valid and that organizations retain the competence, authority, and capacity to monitor systems, exercise oversight, intervene, and discontinue unsafe use.

\mpar{Studying AI Act translation in practice}
Emerging empirical work reports on uncertainty about how to operationalize high-risk requirements and what would be sufficient for compliance~\cite{wagner_ai_2026}, ambiguity and technical tradeoffs in interpreting explainability obligations \cite{dhaini-etal-2025-regulation}, and ``selective integration'' of compliance processes across regulations \cite{restrepo_amariles_compliance_2026}. Co-design with practitioners and compliance experts has also produced an impact assessment template \cite{bogucka_co-designing_2024}. These studies extend research on how practitioners encounter responsible AI in organizations, which has shown that fairness toolkits, checklists, and principles are reshaped by organizational constraints and may struggle to influence consequential organizational decisions~\cite{holstein_improving_2019, madaio_co-designing_2020, rakova_where_2021}.
Participatory approaches have been proposed to bring affected and expert stakeholders into AI policy~\cite{barnett2025envisioning}, with the recognition that participation alone does not resolve design problems~\cite{sloane_participation_2022}.

We complement existing work on the interpretation and operationalization of AI Act requirements by examining how proposed compliance arrangements can fail in practice. Across hypothetical high-risk AI applications, we trace interactions between technical, organizational, and evidentiary requirements and identify the pathways through which formally plausible measures may become ineffective.

\section{Method}
\label{methods}

We conducted a participatory workshop within SIG-AI ACT, a special interest group concerned with translating the EU AI Act into technical requirements and design practices.
The study combined pre-event data collection and a structured full-day workshop with group activities.
We grounded the approach in participatory design and community-based research, which treat researchers and practitioners as co-investigators of a shared problem rather than as informants \cite{barnett2025envisioning, torkamaan_challenges_2024}. The choice of a workshop over interviews or a survey was deliberate to understand where the translation of regulatory obligations stalls and why stakeholders diverge, and to create conditions under which such divergences can be observed, articulated, and examined collectively.

\subsection{Pre-workshop data collection}

During registration, participants were asked to submit questions and concerns regarding the integration of the AI Act into existing workflows. This served two purposes. It provided a baseline account of participants' concerns before any framing by the research team, reducing the risk that workshop discussions would merely reflect the organizers' own preferences, and it informed the design of the breakout sessions.
In total, $46$ registrants contributed questions. The research team inductively coded these responses into thematic categories that covered tensions, bottlenecks, proposed solutions, application challenges, and open questions. This analysis constituted the first empirical layer for answering the research questions.

\subsection{Workshop}
A total of $32$ participants (M = $17$, F = $15$) attended the workshop in person, $22$ from academia and ten from practice. Participants' primary expertise and background are categorized into four groups: technical (n=$18$), legal (n=$6$), social \& ethical (n=$6$), and governance (n=$2$). All participants received an informed consent form before the event and signed a printed copy on the day.

\begin{figure}[h]
    \centering
    \includegraphics[width=0.99\linewidth]{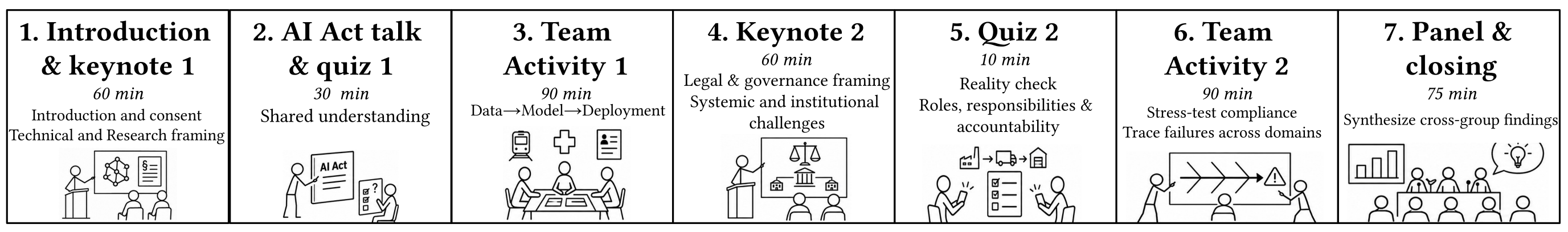}
    \caption{Phases of the full-day workshop}
    \label{fig_methods}
\end{figure}

The workshop\footnote{\url{https://sig-ai-act.github.io/workshop.html}}took place on 6 March, 2026 at Delft University of Technology, the Netherlands, and consisted of seven phases (Figure~\ref{fig_methods}). The morning and afternoon sessions each included a keynote, a quiz, and a structured team activity. Participants were organized into interdisciplinary teams of approximately five to six, each working throughout the day on one of three hypothetical high-risk AI applications: a harassment and violence prevention assistant for public transport (Use Case 1), a multimodal pain estimation system for patients (Use Case 2), and a recruitment assistant (Use Case 3). The scenarios were designed to expose challenges in interpreting and implementing AI Act obligations across different operational contexts. In the morning session, following an introduction to the AI Act and a quiz to establish a shared understanding of relevant provisions, teams completed a worksheet covering three AI lifecycle stages: data collection and processing, model development, and deployment. They identified the system's likely risk classification, relevant actors and provider–deployer responsibilities, applicable regulatory provisions, and the evidence needed to demonstrate compliance. Completed worksheets were collected as primary data.

The afternoon session focused on governance, accountability, and potential compliance failures. Following a keynote and a second quiz, teams revisited their original use cases using an adversarial approach. Through fishbone (Ishikawa) diagrams, participants identified potential causes of non-compliance across six domains: data, model behavior, human oversight, transparency, organizational governance, and post-deployment monitoring. Each team also developed at least one realistic worst-case scenario illustrating how technical or organizational failures could escalate into regulatory non-compliance or harm to fundamental rights. Using the same use cases across both activities enabled comparison between the initial identification of compliance requirements and the subsequent examination of failure pathways. Fishbone diagrams and scenario narratives were collected as research data. The workshop concluded with a panel discussion synthesizing insights across the three applications.

\subsection{Data Analysis}
The study generated five categories of data: (i) pre-workshop registration responses; (ii) response distributions from both quizzes; (iii) morning worksheets documenting role assignments, regulatory provisions, and proposed compliance evidence; (iv) afternoon fishbone diagrams and worst-case scenario narratives; and (v) panel discussion notes. Group discussions were documented by designated scribes, and completed worksheets were transcribed following the workshop. We used reflexive thematic analysis \cite{braun2019reflecting} to identify recurring challenges in translating AI Act obligations into technical requirements, organizational practices, and compliance evidence. Pre-workshop responses were analyzed separately to identify participants' initial concerns and inform the workshop design. Workshop materials were then analyzed separately for each use case through two coding passes, integrating codes from the different data sources into case-level thematic clusters. These clusters were subsequently compared across the three use cases to identify recurring patterns and context-specific differences. This process informed the six themes presented in Section~\ref{workshop_result}.

\pagebreak

\section{Results}

\subsection{Pre-workshop inputs: what practitioners and researchers bring to the translation problem}

From the pre-workshop corpus, we developed three recurring themes: structural tensions that registrants identified in attempts to apply the Act within current practice, operational bottlenecks that block specific compliance steps, and application challenges tied to particular technical capabilities or domains.

\subsubsection{Structural tensions}
\begin{figure}[h]
    \centering
    \includegraphics[width=0.8\linewidth,trim=0 20mm 0 0, clip]{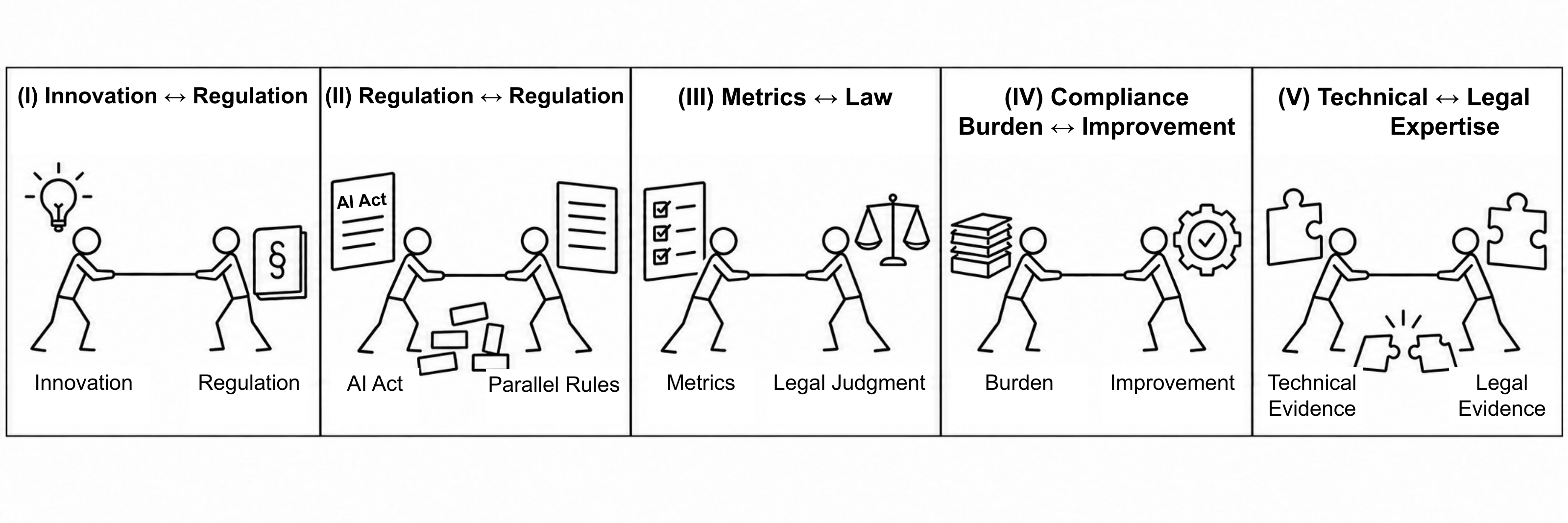}
    \vspace{-10pt}
    \caption{Structural Tensions in Translating AI Regulation into Practice}
    \label{fig_structural_tention}
\end{figure}

Registrants identified several structural tensions in translating the AI Act into practice (Fig.~\ref{fig_structural_tention}). One concerned the relationship between innovation and regulation, with participants expressing concerns that compliance requirements could constrain system design, delay the development of socially beneficial applications, and limit innovation. A related tension emerged between regulatory principles and their practical implementation. Although the Act establishes general obligations, registrants highlighted the difficulty of translating these into concrete requirements for specific AI systems, organizations, and workflows.

A second tension concerned the interaction between the AI Act and other EU regulatory initiatives. Registrants identified the digital omnibus simplification package and uncertainty surrounding harmonized standards as challenges for compliance planning. Their concern extended beyond identifying applicable requirements to understanding how organizations could plan and demonstrate compliance while the regulatory framework and supporting standards continued to evolve.

A third tension emerged between technical metrics and legal requirements. Registrants questioned whether fairness metrics, robustness benchmarks, and accuracy thresholds could adequately represent legal concepts such as non-discrimination, proportionality, and meaningful human oversight. They highlighted the limitations of using technical measurements to support legal judgments, particularly when normative concepts \say{cannot be fully operationalized without loss of meaning}.

A fourth tension concerned whether compliance requirements should be viewed primarily as an administrative burden or as an opportunity for improvement. While some registrants questioned whether additional compliance obligations would meaningfully improve safety, others considered regulation a potential driver of better AI development and governance practices. 

Finally, registrants highlighted the challenges of collaboration between technical and legal experts. These challenges extended beyond communication, reflecting differences in how disciplines interpret requirements, evaluate evidence, and assign responsibility. Integrating these perspectives into a coherent sociotechnical assessment was therefore seen as requiring more than simply bringing experts together.

\subsubsection{Operational bottlenecks}
\label{sec_operational_bottleneck}

Registrants identified a set of specific procedural and institutional bottlenecks where compliance work stalls concretely. Certification was the most frequently named: registrants described uncertainty about conformity assessment pathways, particularly in the absence of harmonized standards, as a practical blocker rather than an abstract concern.
Enforcement was a second bottleneck, in a different register: registrants were uncertain not only about what enforcement would look like in practice, but about who would have the authority, capacity, and technical expertise to conduct it meaningfully. They warned that when compliance criteria are unclear, organizations may optimize for what looks compliant to an auditor rather than for substantive safety or rights protection.
Two further bottlenecks were technical. Best practices for auditing datasets, replaying decisions, and demonstrating operational accountability were described as absent or nascent. And, reproducible decision execution---the ability to reconstruct why a system produced a particular output in a particular context---was identified as a capability that many current AI pipelines do not natively support.

\subsubsection{Application-specific challenges}
Unlike the cross-cutting procedural and institutional bottlenecks above, registrants also identified challenges arising from particular AI capabilities, use cases, or classification boundaries. Deepfakes and synthetic content were cited as examples where the Act's disclosure and marking obligations may be technically difficult to implement and enforce. Registrants also raised risk classification as an application-specific challenge, particularly for systems that sit at the edge of Annex III categories, where determining the applicable risk category and associated transparency obligations may be unclear.
They also highlighted the difficulty of reconciling different interpretations of the AI Act across different professional communities, legal, technical, and operational, framing this as both a coordination and governance challenge. They noted that such disagreements are especially consequential in high-stakes domains such as safety-critical infrastructure.

\subsection{Workshop inputs: where the translation problem stalls in practice}
\label{workshop_result}
From participants' contributions across the healthcare, hiring, and transport use cases, we developed six themes. They concern (i) what systems measure and what their outputs establish, (ii) how oversight and responsibility are organized, and (iii) how changes in operating conditions and use affect compliance. Quotations from worksheets and scribe records illustrate the analysis.

\subsubsection{Bias can emerge from data, design choices, and deployment}
\label{sec_datarep}
Participants identified several ways in which bias could arise from the relationship between data representation, measurement choices, and deployment conditions. In \emph{healthcare}, subjective experiences such as pain had to be inferred from indirect signals that might not adequately represent all patients. In \emph{hiring}, historical records could reproduce earlier organizational decisions, including \say{bias embedded in 10 years of internal hiring data}, while access to sensitive attributes needed to assess discrimination was constrained. In \emph{transport}, data coverage varied across locations, populations, languages, seasons, and events. Across the three cases, participants emphasized the need for representative datasets, data-quality documentation, and subgroup performance assessments. Although synthetic data were proposed as a possible response to limited patient data, participants recognized that increasing data availability would not necessarily resolve problems with how the intended construct was measured.

A second source of bias concerned the selection of features and fairness criteria. In \emph{hiring}, participants highlighted that competing fairness objectives could produce different outcomes, particularly when the number of available positions was limited. One group concluded that \say{unbiased hiring is impossible}, reflecting the difficulty of satisfying multiple fairness criteria simultaneously. In \emph{transport}, characteristics such as accent, appearance, location, and incident history could function as discriminatory proxies even when sensitive attributes were not explicitly included. These examples illustrate how bias may arise not only from inadequate representation but also from decisions about which attributes and evaluation criteria are considered relevant.

Finally, participants identified feedback mechanisms through which system deployment could introduce or reinforce bias over time. In \emph{healthcare}, treatment decisions could alter the pain signals subsequently observed. In \emph{hiring}, participants noted that \say{output from model influences training dataset}: earlier selections could shape future training data, while candidates adapting their applications to algorithmic criteria could introduce \say{gamed CVs}. In \emph{transport}, increased monitoring of particular locations could produce more recorded incidents from those areas, potentially reinforcing their classification as high-risk. These examples suggest that bias is not solely a property of the initial training dataset but may emerge through interactions between system outputs, human behavior, and subsequent data collection. We therefore interpret participants' concerns as identifying three interconnected challenges for data governance: the validity of data representation, the justification of fairness criteria and proxies, and the monitoring of feedback effects after deployment.

\subsubsection{A Human in the loop does not ensure effective oversight}
\label{sec_oversight}
Participants identified different ways in which formal human oversight could fail in practice. In \emph{hiring}, the timing of human review was critical: intervention after automated filtering or ranking could come too late to prevent unfair exclusions. Participants also noted that limited expertise, time pressure, deskilling, and requirements to justify deviations from AI recommendations could discourage independent judgment. In \emph{transport}, real-time alerts left little time for verification, while labels such as \say{assault likely} could convey greater certainty than the evidence warranted. Participants also questioned what would constitute sufficient grounds to override an alert. In \emph{healthcare}, participants described a chain of deference in which physicians delegated decisions to nurses, nurses assumed that AI-generated scores had already been validated, and the system became the default basis for judgment. Across these cases, AI systems could remain formally advisory while becoming functionally determinative, creating risks of unfair exclusion, inappropriate intervention, or clinical harm.

Participants in the transport group challenged individualized explanations of automation bias, arguing \say{let us make it less an individual character problem and more a systemic problem}. They linked uncritical reliance on AI recommendations to workload, resource constraints, and organizational pressure. Under time pressure and with limited information, operators could struggle to verify or override system outputs. As one participant explained, \say{it is easier to say: whatever the system says, because I do not have all the information.} In \emph{healthcare}, participants also warned that overreliance could develop gradually through repeated successful interactions---\say{the system is always right \ldots{} I don't need to call the specialist}---and might therefore remain undetected in short-term evaluations. Together, these accounts suggest that effective human oversight is an organizational capability rather than simply a designated role or interface feature. It requires clear decision-making authority, sufficient information, competence and time, workable escalation and override procedures, organizational support for challenging automated recommendations, and ongoing monitoring of how people rely on AI systems in practice.

\subsubsection{Responsibility is distributed differently from the capacity to discharge it}\label{sec_ownership}
Across the three use cases, participants described responsibility as fragmented across organizations, technical components, and stages of the AI lifecycle, with no single actor controlling the complete system. Mapping actors to regulatory and professional responsibilities revealed that providers and vendors controlled model development, updates, and technical documentation; deployers controlled procurement, local data, workflow integration, and how outputs informed decisions; frontline professionals acted on those outputs; and applicants, patients, and passengers experienced their consequences. Responsibility was therefore distributed differently from the information, technical capabilities, and authority needed to discharge it.
 
In \emph{hiring}, employers often procured proprietary third-party systems in which \say{access is refused to audit}, with no knowledge transfer and no contractual provisions for maintenance, incidents, appeals, or system retirement; the party able to make the system compliant sat outside the organization accountable for it. In \emph{healthcare}, providers, hospitals, integrators, and clinicians controlled different parts of the clinical system, and participants described responsibility between deployer and provider as undefined---\say{no clear responsibilities between deployer and provider}---which complicated incident reconstruction, monitoring, and withdrawal. In \emph{transport}, responsibility depended on who determined the system's actual use: \say{it is about who is the actor deciding to put it into practice}. Participants held that the \say{public vs private arrangement must be defined first}, before any technical assessment, and that arrangements among vendors, private operators, municipalities, infrastructure owners, and police became particularly unclear when a bounded safety system was repurposed for surveillance or law enforcement (Section~\ref{sec_scope}).

Across the cases, accountability weakened when responsibility, evidence, and authority were separated. The actor observing harm might lack technical logs, while the actor controlling the model might lack access to operational outcomes. Documentation was therefore treated as effective only when the actor who needed it could access, interpret, and act on it. 

\subsubsection{Changing conditions can invalidate prior compliance evidence}
\label{sec_continuedop}
Systems can outlive the conditions that made them compliant: they may remain technically functional while their operating environment, their data, or the organization that supports them changes.

The first source of uncertainty about continued compliance was the environment. Beyond the feedback effects described in Section~\ref{sec_datarep}, participants pointed to shifts the system did not cause: seasonal changes and special events in transport; new patient populations, new sensor hardware, and changing staff behavior in health; and changing labor markets in hiring, with COVID cited as an example of historical hiring data becoming unsuitable. Evidence of representativeness and performance established at validation could therefore become obsolete without any change to the system itself.

A second source of lost compliance was organizational. Developers could leave once a system reached the less attractive maintenance phase, or the vendor could cease operating while deployers continued to use the tool: \say{the company that built it may be bankrupt by the time monitoring matters.} The deployer might then retain responsibility without the expertise, documentation, or technical access needed to fulfill it (Section~\ref{sec_ownership}). In \emph{healthcare}, participants were also uncertain when a hospital that adapted a system, or updated it with live patient data, would take on responsibilities associated with a provider; updating could invalidate the assumptions of the initial validation and create new obligations concerning privacy, data provenance, and provider--deployer roles. In \emph{hiring}, the system could be difficult to modify at all because of path dependence and limited access to sensitive data, and participants asked how much retraining constituted a change.

Participants consequently called for explicit retraining triggers, auditing of updates, and \say{system sunsetting clauses}: when the conditions necessary to reassess or update a model cannot be sustained, continued deployment should not become the default. The common concern was therefore not model aging as such, but the growing mismatch between a continuing system and the conditions under which its use was originally justified.

\subsubsection{AI systems can perform well yet fail to support their intended decisions}
\label{sec_fit}
Across the three use cases, participants questioned whether technical performance was sufficient to establish that an AI system was appropriate for the decisions it was intended to support. In \emph{healthcare}, participants distinguished between optimizing pain reduction and achieving broader clinical goals, such as patient well-being, quality of life, and avoiding unnecessary medication. A system could accurately estimate pain while supporting interventions that clinicians considered inappropriate. Participants questioned whether the objective was \say{reducing pain, or is it well-being?}, with one describing the risk as \say{the wrong optimization problem}. They also raised concerns about how the suitability and safety of such systems could be established without evaluation under actual clinical conditions. These discussions highlighted a potential mismatch between the system's optimization objective and the practical goals of the decision it was designed to support.

In \emph{hiring}, participants identified a related gap between technical evaluation and legal or organizational judgments. They questioned whether computational measures of bias or fairness necessarily constituted discrimination---\say{does bias automatically lead to discrimination?}---and described \say{a mismatch between what we are measuring versus what we want to}. Fairness metrics and audits could provide useful evidence, but participants questioned whether these adequately captured the treatment of individual applicants, particularly when evaluating proprietary systems. An acceptable aggregate fairness score could conceal problematic decisions, while a statistical disparity would not necessarily establish unlawful discrimination. In \emph{transport}, participants similarly questioned whether algorithmic risk assessments were \say{suitable for the task}, particularly when model outputs differed from operators' contextual judgment. They debated whether professional caution, intuition, and risk preferences could or should be represented in the system's objective. Across these cases, participants distinguished between measuring a technical property and demonstrating that the property was relevant to the decision being made.

Transparency raised a further question of alignment between the information provided and the needs of different stakeholders. Participants asked not simply how much information should be disclosed, but to whom, for what purpose, and in support of which decision. In \emph{hiring}, explanations intended for HR staff did not necessarily enable applicants to understand or contest decisions; participants described such explanations as \say{built for the wrong stakeholder}. In \emph{healthcare}, participants questioned whether communicating high pain or distress scores directly to patients might be \say{counter-productive}, while also raising concerns about the faithfulness of post-hoc explanations. In \emph{transport}, participants argued that signage might formally disclose AI-enabled surveillance without meaningfully informing passengers of changes in how existing cameras were being used. They also questioned how passengers could challenge or correct an incorrect classification. Taken together, these findings suggest that metrics, predictions, and explanations become meaningful evidence only in relation to the actors, decisions, and operational contexts they are intended to support. Technical validity alone does not establish decision suitability or regulatory compliance.

\subsubsection{AI systems can be repurposed without reassessing their risks}
\label{sec_scope}

Across the three use cases, participants raised concerns about systems gradually being used beyond the boundaries under which they were designed and assessed, and the cases illustrated different routes by which this could happen. The transport scenario itself posed the possibility of expansion towards policing; participants in the other two groups raised comparable concerns without being prompted.

In \emph{transport}, a bounded harassment-detection system could expand into broader surveillance or policing through new uses of camera and audio streams or police access---\say{misuse for mass surveillance}. Participants noted that adding an AI layer to existing cameras did not merely add a component: \say{by changing the intended use of the system, you change the system as a whole}. They discussed \say{function creep}, \say{ring-fencing}, and limiting the system to its original purpose, and asked when the change point arrives at which a deployer's duties change. In \emph{healthcare}, a pain estimate intended to inform care could become increasingly tied to nurse visits, medication decisions, or even automated delivery---\say{connecting the system to the IV pumps}---so that a monitor became an actuator; this is the operational counterpart of the deference described in Section~\ref{sec_oversight}. In \emph{hiring}, candidate data could be repurposed beyond their stated purpose, and the system's risk depended on whether it supported, automated, or effectively determined selection: \say{cutting 200 applicants to 50 is not the same as choosing one of three finalists}.

What moved beyond the assessed boundary therefore differed across cases: the purpose, the data use, who had access, or how much operational authority the output carried. The shared difficulty was identifying when such a gradual change became a materially different use requiring renewed assessment or a new allocation of responsibility. This differs from the environmental and organizational drift described in Section~\ref{sec_continuedop}: a system may become outdated without changing what it is used for, whereas here the use changes without a corresponding reassessment of risks, safeguards, and responsibilities.

Across the themes, participants also proposed responses: retraining triggers, sunsetting clauses, explicit specification of who decides what and when, and limits on intended purpose. Section~\ref{sec_discussion} considers how these findings relate to the tensions identified in the pre-workshop intake and their implications for implementing the Act.

\section{Discussion}
\label{sec_discussion}

Across the six themes, a central challenge emerged: translating regulatory obligations into practice requires accounting for AI systems whose purposes, operating conditions, and responsibilities are distributed across actors and evolve over time. The pre-workshop tensions (Figure~\ref{fig_structural_tention}) and quiz results suggest that these difficulties extend beyond familiarity with the Act's provisions. Although harmonized standards and regulatory guidance can reduce uncertainty, their application still requires context-specific judgments. We identify three interrelated tensions: between technical metrics and legal concepts, between documented compliance and organizational capacity, and between initial conformity assessments and changing deployment conditions. These tensions structure the following discussion around the three research questions.

\subsection{Technical translation: aligning system objectives with their intended use} \label{sec_disc_measure}

Participants treated representativeness not merely as a matter of data quality, demographic coverage, or subgroup performance, but as dependent on the construct being measured, the context of use, and changes resulting from interventions and system updates. Article~10 reflects this perspective by requiring data to have statistical properties appropriate to the intended purpose and setting. This presupposes a clear specification of what the data are intended to represent. Representativeness therefore cannot be assessed through demographic coverage alone, just as favorable fairness metrics cannot, by themselves, establish the absence of unlawful discrimination. 

The technical difficulty follows from the above observation. Before representativeness and bias can be meaningfully measured or addressed, they must first be defined in relation to the specific problem and context of use. 
However, defining these concepts requires determining what the system should measure and optimize, which trade-offs are acceptable, what constitutes bias in the specific decision context, and how these considerations should be revisited as the system evolves. Participants found these choices contested across all three use cases, highlighting that defining and assessing representativeness and bias requires an interdisciplinary, human-centered process that extends beyond technical validation. The reason for that is that a system's own operation changes what is later observed (Section~\ref{sec_datarep}), so the definitions and the evaluations built on them have to be revisited throughout the lifecycle. Two consequences follow: the work may fall outside the skills of a development team, and its outcomes are use-case specific, which limits how far any one operationalization can generalize.

Technical metrics and tests provide evidence of particular system properties, but their relevance to compliance depends on how those properties relate to the system's intended purpose and associated risks. The AI Act already addresses this relationship: Article~9 requires continuous risk management, Article~9(8) requires testing against metrics appropriate to the intended purpose, Article~15 establishes accuracy and robustness requirements, and Annex~IV(2)(b) requires documentation of optimization objectives and key design choices. However, these provisions do not eliminate the need to justify which objectives should be prioritized, how competing values are reconciled, and whether the selected metrics adequately capture the risks of a particular application. In \emph{healthcare}, accurate pain estimation may conflict with broader goals such as patient well-being or limiting medication, while in \emph{hiring}, faithfully applying employer-defined criteria may reproduce historical inequalities. 
Our findings therefore emphasize that favorable performance metrics alone do not establish compliance. Instead, the relationship between regulatory objectives, computational targets, and technical assessments must be made explicit and open to scrutiny.

\subsection{Sociotechnical translation: compliance depends on situated organizational capacity}

Our findings (Section~\ref{sec_oversight}) highlight the challenges of integrating AI Act oversight requirements into human, organizational, and deployment practices.
Effective
oversight depends on the relevant domain and technical competence, information about the system
limitations, sufficient time, and the authority and organizational support to challenge the output. It also
depends on workflow and interface design, including where review occurred and whether uncertainty,
escalation, and override options were actionable.
These findings point to a coordination problem in the Act's allocation of oversight responsibilities. Article~14(4) sets requirements for enabling human oversight, while Article~26(2) requires deployers to assign it to appropriately qualified and supported persons. Under Article~14(3)(b), the provider identifies oversight measures for implementation by the deployer. Yet the provider may have limited knowledge of the specific workflow into which the system will be embedded, while the deployer controls staffing, workload, and the consequences of disagreement but may have limited influence over the system's design.
This division creates a risk that oversight measures are technically available but difficult to exercise in practice. Our findings, therefore, suggest that implementing these provisions requires an assessment of how oversight measures interact with actual working conditions, including weaknesses in organizational procedures that may predate the system.

Conformity assessment can establish that a system was designed to be overseeable. Whether it is overseen can only be shown in operation.
The Act assigns oversight and monitoring duties to providers and deployers (Arts. 14, 26(2), 26(5), and 72). Our findings highlight the difficulty of using monitoring evidence to establish whether oversight remains effective in practice.
Override frequency alone does not show whether people exercise sound judgment. Few overrides may reflect appropriate reliance or uncritical acceptance, while frequent overrides may be poorly justified. Assessing oversight, therefore, requires evidence that people remain able to critically evaluate outputs and intervene as workloads, reliance on the system, and working conditions change.

The mismatch between formal responsibility and practical capacity extends beyond human oversight. Participants described how deployers could remain accountable for system use while depending on providers for technical information, audit access, updates, and maintenance. These dependencies should be addressed during procurement, when contractual arrangements can establish access to documentation, monitoring information, and procedures for intervention or withdrawal. Although Article~25(4) establishes information-sharing requirements between providers and third-party suppliers, it does not necessarily resolve dependencies between providers and deployers. Responsibilities may also change during deployment: under Article~25(1), substantial modification of a high-risk system, or a change of intended purpose that brings a previously non-high-risk system into the high-risk category, can transfer provider obligations to the modifying actor. By contrast, a vendor ceasing operations does not itself transfer provider status, although it may leave deployers without the technical support needed to maintain compliance. These findings suggest that accountability requires more than assigning legal obligations; it depends on ensuring that responsible actors retain the information, authority, and organizational capacity to fulfill them throughout the system lifecycle.

The Act ties reassessment to changes in the system: Article~43(4) requires a new conformity assessment after a substantial modification, while Article 72 requires post-market monitoring to evaluate continuous compliance. Our findings highlight the practical difficulty of identifying when changing deployment conditions undermines the evidence supporting continued use. Changes in use introduce a further question: who recognizes that a system is being relied on for a different purpose, and who has the authority and capacity to respond? Monitoring, therefore, needs to connect changes in populations, working conditions, and actual use to explicit decisions about reassessment, restriction, or retirement. The unresolved issue is how gradual change becomes visible and when it warrants intervention, including whether it amounts to a substantial modification or changes the allocation of provider responsibilities.

\subsection{Legal–evidentiary translation: justifying compliance claims}
Producing technical documentation or demonstrating a system property does not, in itself, establish compliance with a legal obligation. Across the three use cases, participants identified challenges concerning the relationship between technical measurements and legal concepts, the limitations of aggregate assessments, and the continued validity of results under changing conditions. Dataset analyses, model cards, audit reports, performance tests, and records of human oversight may support compliance claims, but their legal significance depends on what they establish and whether this is sufficient for the relevant obligation. For example, in \emph{hiring}, a favorable fairness metric may indicate limited disparities between groups without establishing the absence of unlawful discrimination, particularly when aggregate results conceal the treatment of individual applicants. Similarly, documenting an override mechanism does not necessarily demonstrate meaningful human oversight under Article~14. The legal--evidentiary challenge is therefore to justify how technical assessments support specific regulatory obligations, rather than treating the existence of documentation or favorable performance results as sufficient evidence of compliance.

This challenge is further complicated by interactions between legal obligations. Under Article~10, assessing bias may require processing sensitive personal data, subject to applicable data protection rules and safeguards. Transparency under Article~13 must likewise be reconciled with confidentiality and other legitimate restrictions on disclosure, while documenting human oversight may raise privacy concerns for workers and affected individuals. Grant and Wischik \cite{grant2020show} illustrate this tension by showing how validating explanations of automated decisions may require access to information protected by privacy law. These examples demonstrate that compliance cannot be reduced to maximizing documentation or disclosure. Rather, organizations must justify why particular assessments adequately support specific legal claims while respecting other applicable obligations. Such justifications may be scrutinized through conformity assessment, regulatory supervision, or judicial proceedings, depending on the applicable legal framework.

\subsection{Implications for policy}

Our findings suggest that regulatory guidance and harmonized standards should address not only technical requirements and documentation, but also the conditions under which regulatory safeguards can function in practice. This is consistent with broader research emphasizing the importance of human-centered approaches to integrating AI into sociotechnical systems \cite{torkamaan_challenges_2024}. Policymakers and standardization bodies should provide clearer guidance on how technical metrics relate to legal and professional objectives, and how meaningful human oversight can be assessed under realistic working conditions, including organizational constraints and the capacity to intervene \cite{lauxInstitutionalisedDistrustHuman2024}. Guidance should also clarify how providers and deployers coordinate responsibilities and when changes in operating conditions, system behavior, or intended use warrant reassessment, building on existing approaches to conformity assessment and post-market monitoring \cite{mokander_conformity_2022}. More broadly, our findings highlight the need for regulatory approaches that consider not only whether compliance has been demonstrated at a particular moment, but whether the conditions necessary for continued compliance can be maintained as AI systems and their contexts of use evolve.

\section{Conclusion}
We examined the challenges of AI Act compliance through a participatory workshop involving 32 researchers and practitioners across three hypothetical high-risk AI applications. Our analysis reveals how technical, organizational, and evidentiary requirements interact, creating failure pathways through which seemingly sound compliance arrangements can become ineffective. For instance, favorable technical performance may not establish suitability for decision-making, formal human oversight may be ineffective, organizational responsibilities may exceed actors' capacity to act, and changing deployment conditions may invalidate earlier assessments. Our key contribution is to show how these challenges interact and reinforce one another, rather than treating them as isolated compliance issues.

Our findings highlight a key requirement for scalable and defensible AI compliance. While AI developers increasingly pursue general-purpose technologies designed for use across domains, regulatory compliance must be established in the specific contexts where these technologies are deployed. Bridging this gap requires domain-specific, interdisciplinary working groups to develop reusable methods linking legal obligations to technical, organizational, and evidentiary practices.

\printbibliography[title={References}]
\pagebreak

\end{document}